\documentclass[a4paper,5p, sort&compress, preprint]{elsarticle}

\usepackage[utf8]{inputenc}
\usepackage{graphicx}
\usepackage{float}
\usepackage{hyperref} 
\usepackage{isotope}
\usepackage{upgreek}
\usepackage{sidecap}
\usepackage{siunitx}

\begin{document}

\begin{frontmatter}

\title{\begin{center} The COBRA demonstrator at the LNGS underground laboratory \end{center}   \vspace{0.5cm} \begin{normalsize}The COBRA collaboration\end{normalsize}\vspace{-0.5cm}} 

\author[Hamburg]{J.~Ebert} %\ead{joachim.ebert@desy.de}
\author[Dresden,Dresden-Minnesota]{M.~Fritts} %\ead{matthew\_christopher.fritts@tu-dresden.de}
\author[Dresden]{D.~Gehre} %\ead{daniel.gehre@tu-dresden.de}
\author[Dortmund]{C.~G\"o\ss{}ling} %\ead{claus.goessling@tu-dortmund.de}
\author[Dresden]{T.~G\"opfert} %\ead{thomas.goepfert@tu-dresden.de}
\author[Hamburg]{C.~Hagner} %\ead{caren.hagner@desy.de}
\author[Hamburg]{N.~Heidrich} %\ead{nadine.heidrich@desy.de}
\author[Dortmund]{R.~Klingenberg} %\ead{reiner.Klingenberg@tu-dortmund.de}
\author[Dortmund]{T.~K\"ottig} %\ead{tobias.koettig@tu-dortmund.de}
\author[Dortmund]{K.~Kr\"oninger} %\ead{kevin.kroeninger@cern.ch}
\author[Erlangen]{T.~Michel} %\ead{Thilo.Michel@physik.uni-erlangen.de}
\author[Dortmund]{T.~Neddermann} %\ead{till.neddermann@tu-dortmund.de}
\author[Dortmund]{C.~Nitsch} %\ead{christian.nitsch@tu-dortmund.de}
\author[Hamburg]{C.~Oldorf} %\ead{christian.oldorf@desy.de}
\author[Dortmund]{T.~Quante} %\ead{thomas.quante@tu-dortmund.de}
\author[Dortmund]{S.~Rajek} %\ead{silke.rajek@tu-dortmund.de}
\author[Dresden]{H.~Rebber} %\ead{henning.rebber@desy.de}
\author[Dresden]{O.~Reinecke} %\ead{oscar.reinecke@tu-dresden.de}
\author[Dresden]{K.~Rohatsch} %\ead{katja.rohatsch@tu-dresden.de}
\author[Dortmund,Dortmund-Muenchen]{O.~Schulz} %\ead{oschulz@tu-dortmund.de.de}
\author[Dresden]{A.~S\"orensen} %\ead{arndt.soerensen@tu-dresden.de}
\author[Prague]{I.~Stekl} %{Ivan.Stekl@utef.cvut.cz}
\author[Dortmund]{J.~Tebr\"ugge\corref{cor1}} \ead{jan.tebruegge@tu-dortmund.de}
\author[Dortmund]{R.~Temminghoff} %\ead{robert.Temminghoff@tu-dortmund.de}
\author[Dortmund]{R.~Theinert} %\ead{robert.theinert@tu-dortmund.de}
\author[Hamburg]{J.~Timm} %\ead{jan.horst.karl.timm@desy.de}
\author[Dresden]{T.~Wester} %\ead{thomas.wester@tu-dresden.de}
\author[Hamburg]{B.~Wonsak} %\ead{bjoern.soenke.wonsak@desy.de}
\author[Dresden]{S.~Zatschler} %\ead{stefan.zatschler@tu-dresden.de}
\author[Dresden]{K.~Zuber} %\ead{kai.zuber@tu-dresden.de}

\address[Hamburg]{Universit\"at Hamburg, Institut f\"ur Experimentalphysik, Luruper Chaussee 149, 22761~Hamburg, Germany}
\address[Dresden]{Technische Universit\"at Dresden, Institut f\"ur Kern- und Teilchenphysik, Zellescher Weg 19, 01069 Dresden, Germany}
\address[Dresden-Minnesota]{now at University of Minnesota, Minneapolis, United States of America}
\address[Dortmund]{Technische Universit\"at Dortmund, Lehrstuhl f\"ur Experimentelle Physik IV, Otto-Hahn-Str~ 4, 44227 Dortmund, Germany}
\address[Erlangen]{Universität Erlangen, Erwin-Rommel-Str. 1, 91058 Erlangen, Germany}
\address[Dortmund-Muenchen]{now at Max Planck Institut f\"ur Physik, M\"unchen, Germany}
\address[Prague]{Czech Technical University Prague, Horská 3a/22, 128 00 Praha 2, Czech Republic} 

\cortext[cor1]{Corresponding author. Tel 49 231 755 3542; fax 49 231 755 3688.}

\date{}

\pdfinfo{%
  /Title    (The COBRA Demonstrator)
  /Author   (Jan Tebr\"ugge)
  /Creator  ()
  /Producer ()
  /Subject  ()
  /Keywords ()
}

\begin{abstract}
The COBRA demonstrator, a prototype for a large-scale experiment searching for neutrinoless double beta-decay, was built at the underground laboratory Laboratori Nazionali del Gran Sasso (LNGS) in Italy.
It consists of an array of 64 monolithic, calorimetric CdZnTe semiconductor detectors with a coplanar-grid design and a total mass of \SI{380}{g}. It is used to investigate the experimental challenges faced when operating CdZnTe detectors in low-background mode, to identify potential background sources and to show the long-term stability of the detectors.
The first data-taking period started in 2011 with a subset of the detectors, while the demonstrator was completed in November 2013. To date, more than \SI{250}{kg\,d} 
of data have been collected.
This paper describes the technical details of the experimental setup and the hardware components.
\end{abstract}


\end{frontmatter}

\section{Introduction}
\begin{figure}
 \includegraphics[width=0.99\columnwidth]{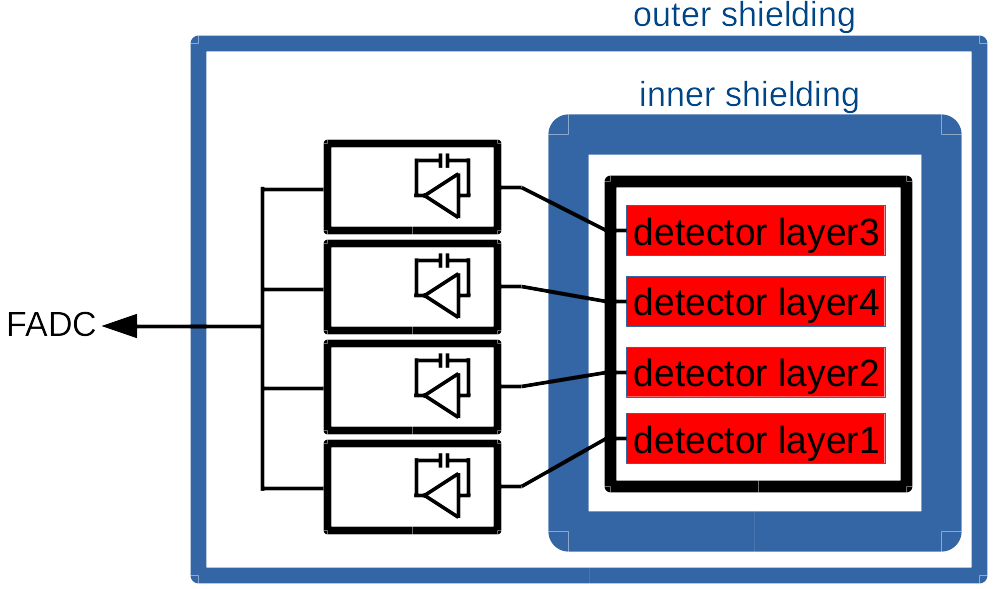}
 \caption[Sketch of the COBRA demonstrator]{Sketch of the COBRA demonstrator. The detector layers three and four are swapped to have the better-performing one in the center of the demonstrator.}
 \label{fig:principal_sketch} 
\end{figure}

The aim of the \underline{C}admium Zinc Telluride \underline{0}-Neutrino Double \underline{B}eta \underline{R}esearch \underline{A}pparatus (COBRA) collaboration \cite{zuber:2001vm} 
is to search for neutrinoless double beta-decay. The employed CdZnTe crystals contain nine isotopes which can undergo double beta-decay of all types. The main focus is on \isotope[116]Cd with a \textit{Q}-value of 2\,814\,keV \cite{Rahaman2011412}, which is above all prominent  
natural gamma lines. 
Neutrinoless double beta-decay can only occur if neutrinos have a non-vanishing rest mass and if they are their own antiparticles, so-called Majorana particles.
In that case, neutrinoless double beta-decay can happen in even-even nuclei if single beta-decay is forbidden or at least strongly suppressed.
In this process no neutrinos are emitted by the nucleus and the energy released in the decay is carried away only by the two electrons. Consequently, the experimental signature in the sum energy spectrum of the electrons is a line at the \textit{Q}-value of the decay. 
Assuming that the decay is found, the effective Majorana mass $\langle m_{\nu_e} \rangle$ can be calculated from the measured half-life of the decay using the relation
\begin{equation}
    \left( T^{0\nu}_{1/2} \right)^{-1} = G^{0\nu}(Q,Z) \, \left| M^{0\nu}_{GT} - M^{0\nu}_F \right| ^2 \left( \frac{\langle m_{\nu_e} \rangle}{m_e} \right)^2 \,, 
    \label{eq:half_live_neutrino_mass}
\end{equation}	
where $G^{0\nu}(Q,Z)$ is the phase-space integral, $Q$ is the decay energy, $Z$ is the proton number of the isotope, $M^{0\nu}_{GT}$ and $ M^{0\nu}_F$ are the Gamow-Teller and Fermi nuclear matrix elements, and $m_e$ is the electron rest mass. 
The calculation of nuclear matrix elements is a severe theoretical nuclear structure problem and
various methods are used to calculate them \cite{Suhonen:1998ck, Caurier:2007qn}.\\ 
Observing neutrinoless double beta-decay would not only allow for calculation of the neutrino mass and reveal the Ma\-jo\-ra\-na-nature of the neutrino, it would also open the door to new physics by being the first documented lepton-number violating process.\\
It is known that double beta-decay is associated with very long half-lives -- more than $10^{25}$\,y \cite{rodejohann}. 
As a result, the expected count rates for experiments searching for neutrinoless double beta-decay are very low. 
Consequently, the most important experimental requirements are to use efficient detectors, accumulate a large number of source atoms  
and reduce the background to a reasonably low level. These requirements drove the design of the COBRA demonstrator: it consists of
an assembly of detectors, a complex shielding structure, a suitable experimental infrastructure and a data acquisition (DAQ) system that allows for pulse-shape analysis. A sketch of the setup is shown in \autoref{fig:principal_sketch}.\\
This paper discusses the technical aspects of the COBRA demonstrator installed at the LNGS underground laboratory.
After an introduction to the COBRA demonstrator in \autoref{sec:cobra}, the shielding concept is described in \autoref{sec:shielding}, followed by a discussion of the CdZnTe detectors  
in \autoref{sec:detectors}. 
After that, the different components of the electronic readout system are presented in \autoref{sec:readout_electronics}. The experimental infrastructure needed to run such an experiment is discussed in \autoref{sec:infrastructure}, \autoref{sec:performance} shows the performance of the demonstrator. Finally, \autoref{sec:summary} summarizes the paper and gives an outlook on future activities.

\section{Overview: The COBRA demonstrator}
\label{sec:cobra}
\begin{figure}
 \includegraphics[width=0.99\columnwidth]{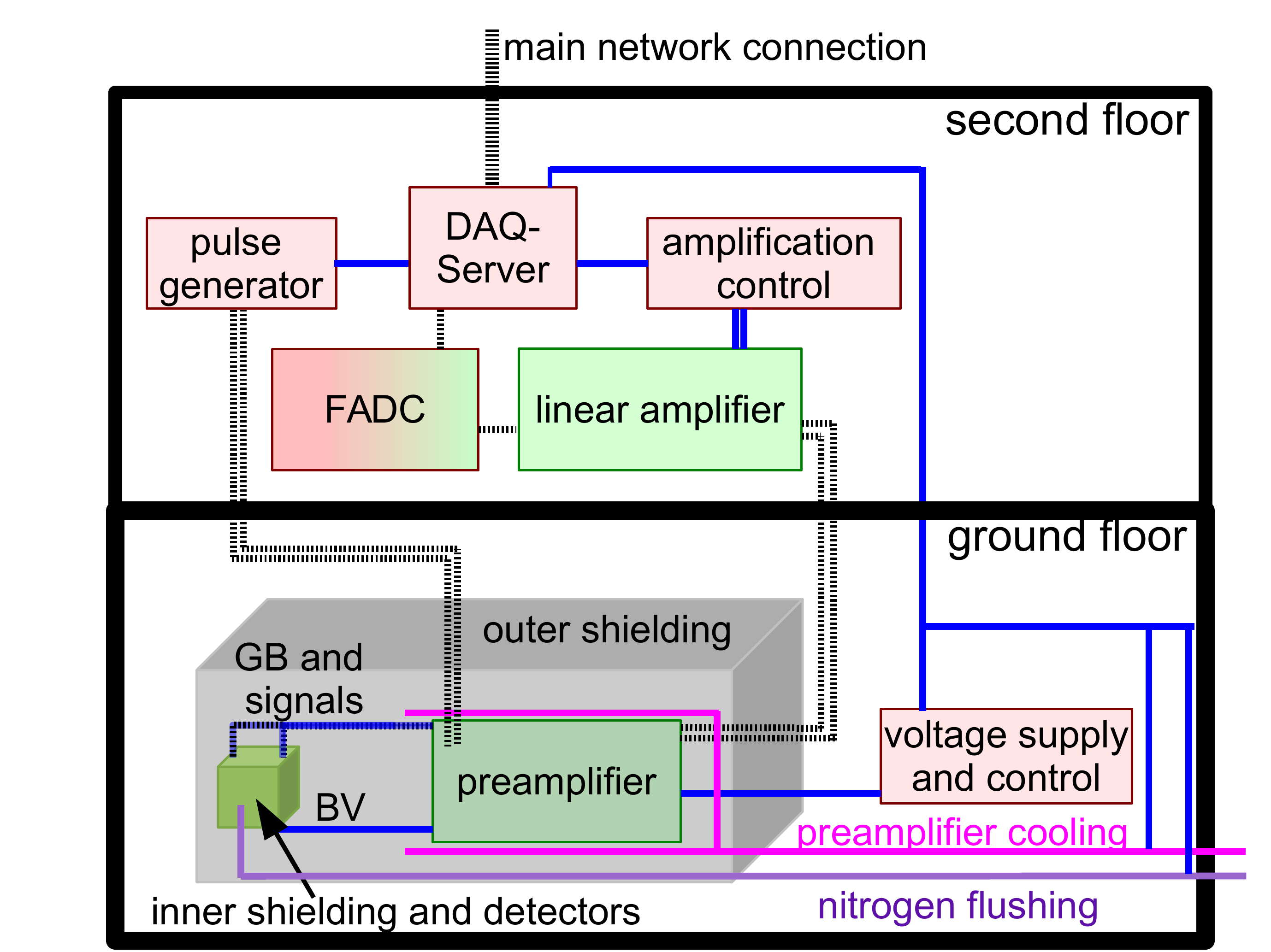}
 \caption[]{Sketch of the general setup of the COBRA experiment, adapted from \cite{tebruegge}. Dotted lines indicate the signal flow, solid ones the supply and control. Double lines symbolize differential signaling. Light and dark boxes show the analog and digital part of the system.}
 \label{fig:general_setup}
\end{figure}

The COBRA demonstrator employs CdZnTe semiconductor detectors with a volume of 1\,cm$^3$. 
An array of 64 of such detectors is currently being operated at the Laboratori Nazionali del Gran Sasso (LNGS) in Italy.
It is covered by about 1\,400\,m of rock which corresponds to a shielding of approximately 3\,800\,m of water equivalent. This reduces the flux of muons from cosmic rays by six orders of magnitude to $(3.41\pm0.01) \cdot10^{-4}\,\text{m}^{-2}\text{s}^{-1}$ \cite{Bellini:2012te}.
The COBRA demonstrator is located between halls A and B in a part of the building that was 
formerly used by the Heidelberg-Moscow experiment \cite{KlapdorKleingrothaus:2000sn}.  
Preliminary studies were done with similar setups prior to 2011, indicating that the operation of CdZnTe detectors should be further investigated \cite{Bloxham:2007aa, Dawson:2009ni, Dawson:2009cc, Goessling:2005jw, Kiel:2003sm}. 
The building COBRA uses consists of two floors. The experimental setup and the first part of the readout electronic system are located on COBRA's part of the ground floor. 
The rest of the electronics and the data acquisition system are  
installed in an air-cooled room on the second floor, since it can dissipate waste heat more effectively.
\autoref{fig:general_setup} shows the detailed design and location of the experiment's components on the two floors.

\section{Shielding}
\label{sec:shielding}

\begin{figure*}
\begin{minipage}[c]{\textwidth} 
\centering 
\includegraphics[width=0.8\textwidth]{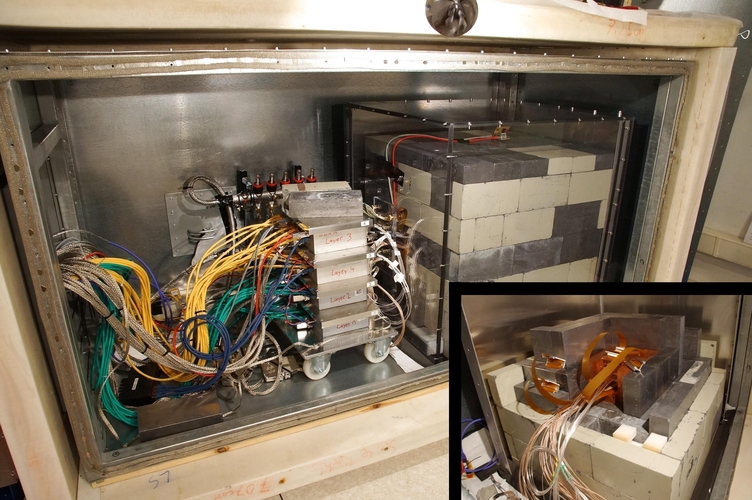}
 \caption[Photograph of the setup]{Photograph of the setup with opened outer shielding. All cables enter the setup from the chute on the left, the stacked preamplifier devices are in the center. On the right side the inner shielding can be seen; the inset shows it partly opened.}
 \label{fig:setup_lower_hut}
\end{minipage}
\end{figure*}

The complex shielding structure used for the COBRA demonstrator consists of an outer and an inner shield. The outer one comprises a neutron shield and a shield against electromagnetic interferences (EMI) with dimensions of approximately $(2\times1\times1)\,\text{m}^3$. 
The neutron shield is made of 7\,cm thick borated polyethylene with 2.7\,wt\% of boron content. Hydrogen-rich plastics, such as polyethylene, moderate neutrons to thermal energies very effectively, while \isotope[10]B has a very high cross section to capture such neutrons 
\cite{muenstermann_2007_phd, neddermann_2014_phd}.
The EMI shield is a construction of iron sheets with a thickness of 2\,mm, which are carefully welded or connected with mesh bands to ensure proper electrical contacts between the different sheets \cite{neddermann_2014_phd}. 
The material was chosen because of its ferromagnetic properties which are highly desirable to shield also against the magnetic component of the electromagnetic interferences. A chute filled with copper granulate acts as a cable feed-through. 
All cables entering the setup pass through this chute in a way that either their outer insulation is stripped off, or an extra metal mesh shielding is added, so that their shielding is connected tightly to the electrical potential of the EMI shield. 
This ensures that none of these cables act as an antenna and transport EMI to the inside of the shielding. 
This is important as a large number of cables are used, for example 32 signal cables.
The functionality of the EMI shield was tested by the independent competence center EMC Test NRW \cite{bernau}.
The chute is also carefully surrounded by a polyethylene construction, so that there are no gaps in the neutron shield.
All preamplifier devices, their cooling plates and the inner shielding are surrounded by the outer shielding.\\
The inner shielding consists of an air-tight sealed housing of metal and polycarbonate plates, which is constantly flushed with evaporated nitrogen creating a slight overpressure. This prevents dust, and especially radon and its decay products, to settle on the detector surfaces. 
Within this housing, the lead shield against radioactive radiation is situated. It is a cube with the side length of $(60\times60\times60)\,\text{cm}^3$ with the detectors in the very center.
The outer part is built from 15\,cm of standard lead, the inner part by 5\,cm of ultra-low-activity lead with a \isotope[210]Pb activity of less than 3\,Bq/kg.

The innermost shielding level (5\,cm) and the support structure of the detectors are made of ultra-pure oxygen-free high-conductivity (OFHC) electroformed copper. 
A total view of the setup including the shielding layers can be seen in \autoref{fig:setup_lower_hut}, the inset shows the partly opened lead shielding.

\section{CdZnTe detectors}
\label{sec:detectors}
\begin{figure}
    \includegraphics[width=.99\columnwidth]{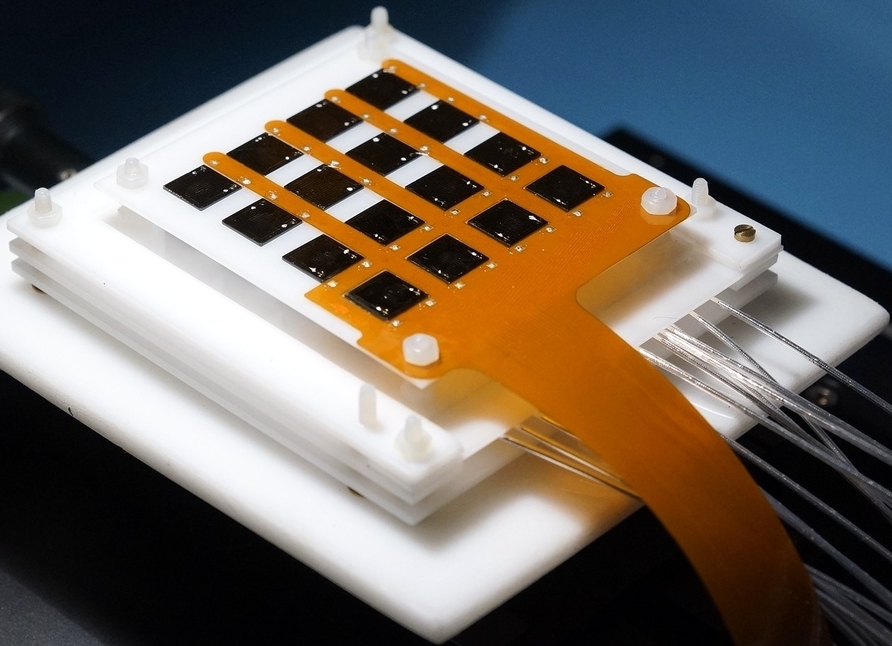}
    \caption[Detector layer]{Photograph of a detector layer holding 16 detectors.}
    \label{fig:layer}
  \end{figure}

\begin{SCfigure}
 \includegraphics[width=0.66\columnwidth]{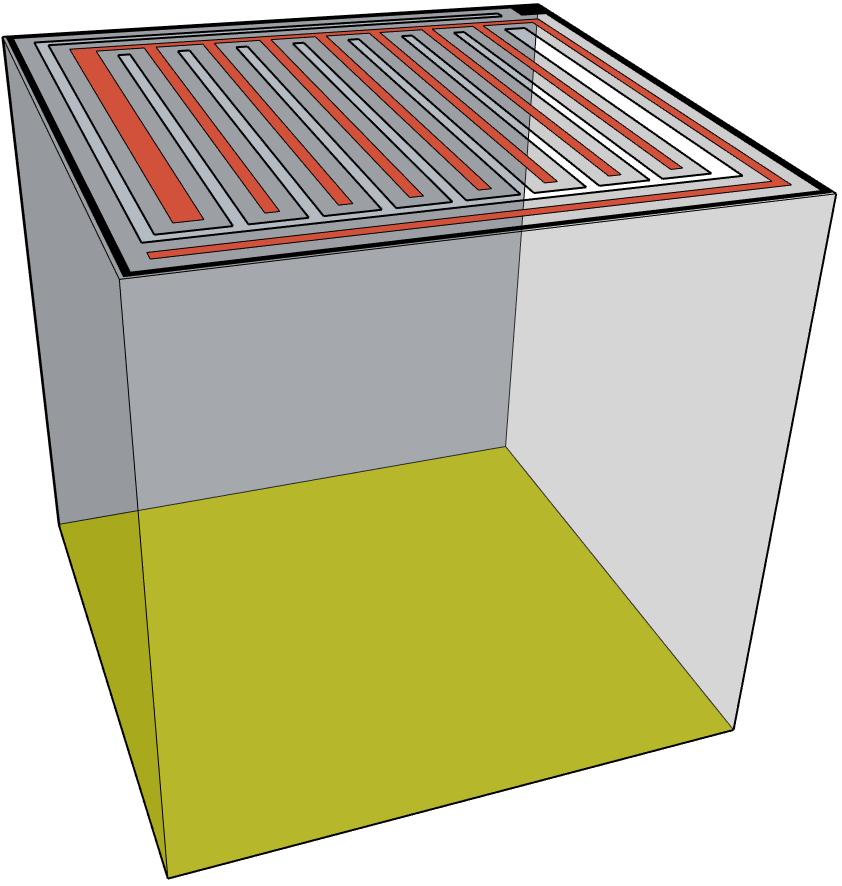}
 \caption[Coplanar grid detector]{Sketch of a coplanar-grid detector. Note the comb-shaped interleaved anode electrodes in red and white, and the planar cathode metalization in yellow on the opposite side. The guard ring shown in black is not instrumented \cite{Fritts:2014qdf}.}
 \label{fig:cpg}
\end{SCfigure}

\begin{figure}
 \includegraphics[width=0.99\columnwidth]{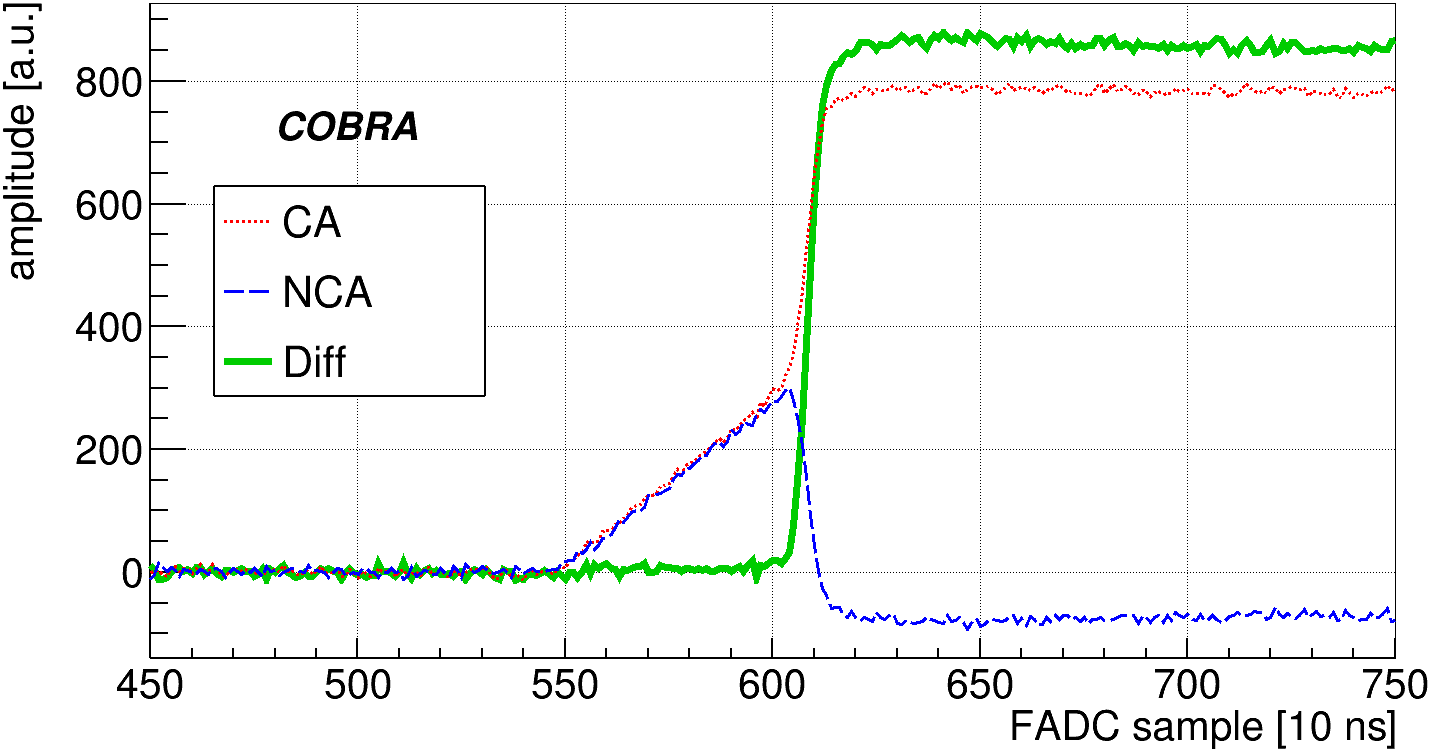}
 \caption[Coplanar grid detector]{Typical pulse shape of the two anode signals (CA and NCA) in dotted and dashed lines, and their difference in solid, which is proportional to the deposited energy.}
 \label{fig:cpg_pulse_shape}
\end{figure}

The COBRA demonstrator uses CdZnTe semiconductor detectors with a coplanar-grid design \cite{cpg94} to mimic a Frisch grid of ionization chambers. CdZnTe can be operated at room temperature because of its high resistivity of approximately $3\cdot10^{10}\,\Omega\, \text{cm}$ and its large band gap of 1.6\,eV \cite{ev_czt_properties}.
64 of these detectors are installed, each with a size of (1$\times$1$\times$1)\,cm$^3$ and a weight of about \SI{5.9}{g}. The whole detector mass is treated as active mass. Effects of analysis cuts are incorporated in the efficiency \cite{Ebert:2015rda}.
16 detectors are mounted on one layer, arranged in a symmetric $4\times4$ structure with a horizontal separation of \SI{4}{\milli\metre}. Two precisely cut Delrin frames provide the mechanical holding system for one layer. Four such layers are arranged on top of each other in the copper support structure with a vertical distance of \SI{1.2}{\centi\metre} in between. One of the layers is shown in \autoref{fig:layer}.
Each detector has one electrical contact on the cathode and two on the anode, referred to as \textit{collecting anode} (CA) and \textit{non-collecting anode} (NCA).
Only those two are read out by the DAQ system; the cathode is not read out. The guard ring surrounding the CA and NCA is not instrumented at all. It reduces distortions in the weighting potential, as it provides the same defined boundary condition for the outermost anode rails. This leads to an improved detector performance \cite{1998NIMPA.411..107H}.
Two different voltages are applied to each detector: the bias voltage (BV) to the cathode and the grid bias (GB) to the NCA, while the CA is kept on ground potential.
The voltages are set individually for each detector to its optimal working point, which were determined with dedicated measurement series \cite{soerensen}. Typical values of the bias voltage and grid bias are \SI{-1.3}{kV} and \SI{-50}{V}, respectively.
\autoref{fig:cpg} shows a sketch of a coplanar-grid detector. A typical example of recorded CA and NCA pulses and their difference, which is proportional to the deposited energy, is shown in \autoref{fig:cpg_pulse_shape}.\\
The electrical contacting of the electrodes is achieved with a \SI{50}{\micro\metre} thin gold wire which is fixed to each of the electrode contact pads with conductive silver lacquer LS200. A glue spot next to it stabilizes it mechanically. 
Studies showed that there is no significant contribution of silver isotopes, especially \isotope[110m]Ag, to the background level \cite{neddermann_2014_phd, hillringhaus}.
Alternative contacting schemes, such as pressure-contacts or soldering, were studied in Ref. \cite{koettig_2008_dipl}, but failed. The gold wires are connected to an RG178 coaxial cable on the cathode side providing the bias voltage. 
For the most recently installed detectors, special low-background high voltage cables are used.\\
The two gold wires on the anode side are glued to a Kapton ribbon cable which provides the constant grid bias voltage and the ground contact, respectively. Furthermore, the anode signals are transmitted on the Kapton cable to the preamplifier devices. 
All cables run on a wavy path through the lead shielding, so that no direct opening through the shielding is pointing towards the detectors.

\section{Readout electronics}
\label{sec:readout_electronics}
The electronic readout system consists of several components: directly outside of the lead shielding, the preamplifiers convert the sensitive detector charge signals into differential voltage signals. This ensures a robust and stable transmission. 
The signals are amplified with the linear amplifiers and converted to single-ended signals. 
The FADCs digitize the signals.\\
The electronic readout system is separated between the two floors of the building. The devices which dissipate a large amount of waste heat, especially the FADCs with approximately \SI{1}{kW}, are located in an air-cooled side room on the second floor. Consequently, the analog part of the readout electronics had to be separated between the two floors. 
As these devices have been custom-made, the issues arising from this, such as a potentially degraded signal quality due to noise and interference, were taken into account completely, i.e. by using differential signal transmission \cite{Schulz, muenstermann_2007_phd, tebruegge}.

\subsection{Preamplifiers}
\label{sec:preamplifiers}
\begin{figure}
 \includegraphics[width=0.99\columnwidth]{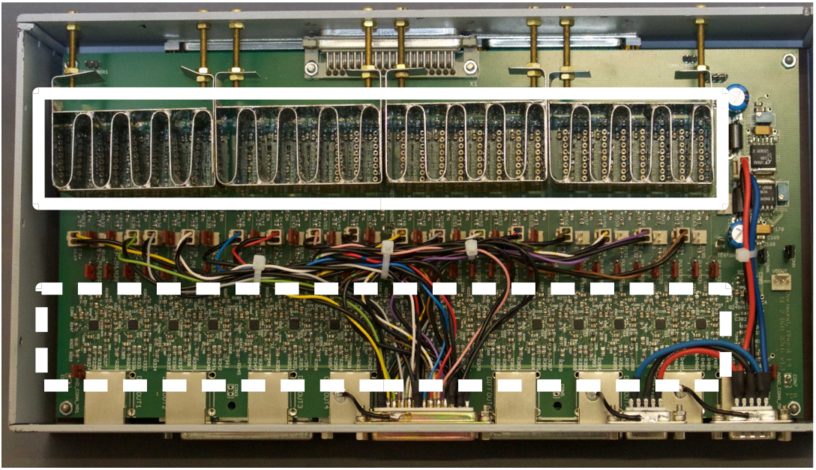}
 \caption[Photograph of the preamplifier]{Photograph of the preamplifier device. The detector signals arrive at the connector on the top side, all 32 channels are parallel to each other from left to right. The metal enclosing for the Cremat CR110 preamplifier components (not inserted) are visible, marked with the white box. In the lower part the conversion to differential signaling is done, indicated with the dashed white box. The distribution of the grid bias voltages is done with the wires coming from the connector on the bottom side.}
 \label{fig:preamp_foto}
\end{figure}

Cremat CR110 charge sensitive preamplifier components convert the charge signals coming from the detectors into voltage signals. These signals 
are transformed to differential signals directly afterwards on the same printed circuit board.\\
One preamplifier device is used for each layer, where each device can handle up to 32 signal inputs (two signal channels for each of the 16 detectors). Each CR110 component is surrounded by a metal shield to suppress crosstalk. 
The differential signal output is fed to RJ-45 connectors, as
matching high-quality differential cables are easily available. Eight ethernet network cables per preamplifier device are needed to transmit all 16 differential detector signals. \\
Separate input lines for signals from a pulse generator are also present for each CR110 component.
The pulse generator signals are supplied via differential signaling to ensure a high signal quality. The conversion to single-ended signals is done directly on the preamplifier printed circuit board.\\ 
Furthermore, the bias voltage and grid bias are also applied via the preamplifier device: the grid bias on the same board, the bias voltage on a separate printed circuit directly beneath the other. The voltages are filtered by an RC low-pass filter.\\
To process all 64 detectors of the COBRA demonstrator, four of such preamplifier devices are needed. These are placed on top of each other next to the inner shielding.
A photograph of the preamplifier device is shown in \autoref{fig:preamp_foto}.

\subsection{Differential signaling}
\label{sec:differential_signaling}
Differential signal transmission is widely used in computer and communication products. It allows for stable and robust transmission as it minimizes crosstalk, electromagnetic interference and noise.
For the COBRA demonstrator, it is mainly used for signal transmission between the preamplifier and linear amplifier, as well as between pulse generator and preamplifier. The length of each cable is \SI{25}{\metre} and covers the passage between the two floors.\\
Standard category 6 network cables are used for the transmission. These are commercially available and are appropriate to transmit sensitive differential signals due to the shielded twisted-pair wires surrounded by an additional shielding layer.
In laboratory tests, cable lengths of \SI{50}{\metre} were tested, even passing areas with enhanced electromagnetic interference, such as housings of running computers and network switches, without a significant deterioration of the signal quality or noise increase.

\subsection{Linear amplifiers}
\label{sec:linear_amplifierts}
The custom-made linear amplifiers are the last element of the analog readout chain. These devices amplify the
detector signals to match the input range of the FADCs, and convert the differential signals to single-ended signals. The amplification can be adjusted in 16 steps of 3\,dB, which is equivalent to a factor of $10^{3/20} = 1.41$
per step. The gain ranges from 0.5 to $0.5\cdot10^{15\cdot3/20} = 89$.
All operations are remotely controllable via an Arduino micro-controller board, acting as an SPI gateway.
The linear amplifiers are operated in 19$''$ NIM crates. Each linear amplifier processes signals of four detectors. It has 16 differential input and eight signal output channels to match the eight input channels of the following FADCs.

\subsection{Flash analog to digital converter}
\label{sec:FADCs}
The flash analog to digital converters (FADCs) are Struck SIS3300 FADC devices. These are VME bus modules with a maximum sampling 
frequency of 100\,MHz and a 12-bit resolution. Additional 3-bit resolution are gained by averaging over 128 samples. Hence, the dynamic resolution is 15 bit \cite{Schulz}.
Two memory banks with $2^{17}$ samples per channel are used. 1\,024 samples are acquired per event, 
resulting in a capacity of 128 events per memory bank. The two-memory-bank mode allows for a dead-time free operation. At 100\,MHz sampling frequency, this results in a timespan of 10.24\,$\mu$s per event that is recorded. As the longest physical signal spans are of the order of 1$\,\mu$s, there is enough pre- and post-trigger information for later off-line analysis methods.\\
Each module has eight input channels. Firmware and hardware were modified to match the specification of the COBRA demonstrator, such as choosing a peak-to-peak input range of \SI{2}{V} and adjusting timing precision.

\section{Experimental infrastructure}
\label{sec:infrastructure}
The experimental infrastructure of the COBRA demonstrator is designed to be remotely controllable. Special care has been taken to include the infrastructure into the shielding concept and the low-background operation.

\subsection{Uninterruptible power supply}
Several uninterruptible power supply (UPS) units are used to ensure stable operation of the experiment, protecting it from rare external voltage breakdowns. Furthermore, these units stabilize and clean the voltage levels.
The UPS units are installed on both floors.
All aspects of them can be controlled remotely. Fully loaded, they can run the experiment for at least 20 minutes if necessary.
Extra UPS power lines without circuit breakers were installed to supply the UPS units because when supplied by the LNGS main line, a high residual current caused occasional power cut-offs. 

\subsection{Voltage supplies}
Different voltages are needed to run the experiment. Therefore several ISEG and WIENER low noise voltage supply devices working in an MPOD module frame are placed close to the experimental setup on the ground floor.\\
These cables are carefully shielded and fed through the chute included in the electromagnetic shielding. Other voltages, such as the supply of the thermal resistor in the nitrogen dewar are less critical. Nevertheless a shielded cable is used here as well to prevent a coupling of distortions to the other voltages. All voltage supply modules can be controlled remotely.

\subsection{Nitrogen-flushing} 
\label{sec:nitrogen_flushing}
\begin{figure}
 \includegraphics[width=0.99\columnwidth]{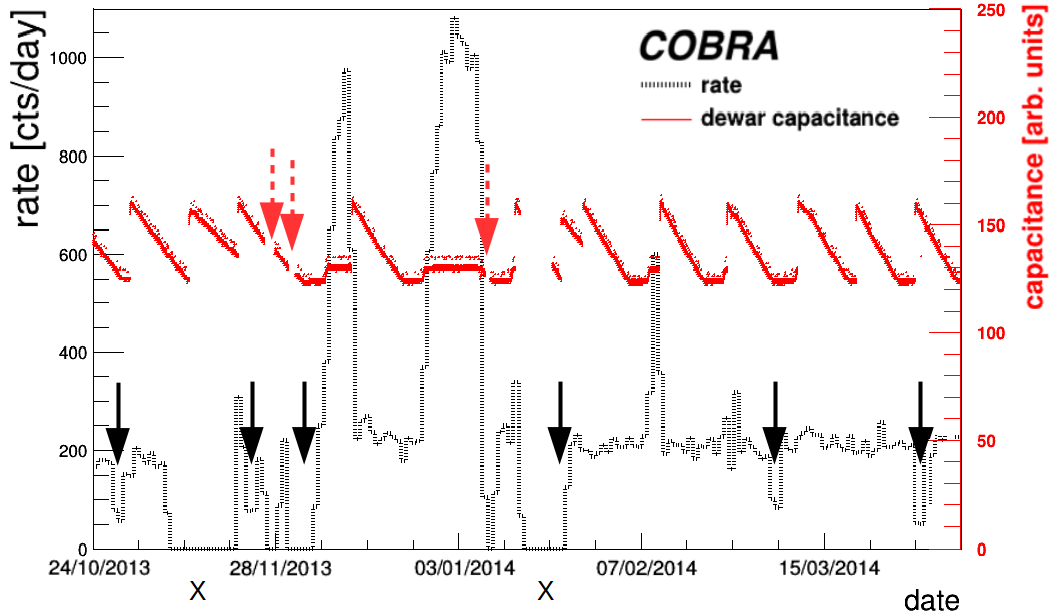}
 \caption[Effect of nitrogen-flushing]{Effect of nitrogen-flushing on the measured data rate of the whole COBRA demonstrator. The solid curve shows the filling level of the nitrogen dewar, for details see text. 
  The rate of events over \SI{500}{keV }surviving data cleaning cuts is shown in dashed. 
  The dashed arrows show power cut-offs due to residual currents, the solid ones calibration measurement periods. The X indicate major upgrade works on the setup. 
  }
 \label{fig:nitrogen_flushing}
\end{figure}

To prevent dust and especially radon settling on the detector surfaces, the inner part of the shielding is constantly flushed with evaporated purified dry nitrogen at a typical rate of 5\,l/min. The dewar of liquid nitrogen is located outside the building. It has a thermal resistor in the liquid phase of 
the nitrogen that can be heated remotely. The nitrogen used for flushing the setup is taken from the gas phase in the dewar, filtered, and fed through tubes into the setup. 
The filtration of the gaseous nitrogen is done by an activated carbon filter which is installed at the bottom of the dewar. 
The filling level is measured with a long pipe-shaped capacitor whose capacitance is directly proportional to the dielectric within (i.e. nitrogen filling level). This capacitor does not reach the bottom of the vessel, because of the activated carbon filter. Consequently, there is still some 
liquid nitrogen left in the vessel, even if the measured capacitance has reached its lowest level. When the liquid nitrogen is used up completely, the temperature of the dewar vessel increases from the temperature of liquid nitrogen to ambient temperature. This will cause the capacitance to increase a little due to thermal effects \cite{muenstermann_2007_phd, alex}.
The effect of the nitrogen-flushing and its rare failure on the measured data rate can be seen in \autoref{fig:nitrogen_flushing}.
The rate is fairly constant over time, but it rises abruptly when the nitrogen-flushing fails. If the flushing is working again, the data rate falls to its lower level quickly as the nitrogen removes radon, its decay products and other contamination from the detectors. 
Another effect of the nitrogen-flushing is a very dry atmosphere of typically 2\% relative humidity, which allows to cool the preamplifier devices without the risk of condensation.
During the upgrade shift in November 2013, the voltage supplies, and hence the nitrogen heating, had been switched off so the liquid nitrogen boiled off slower, which can be seen in a lower slope of the red curve. During the shift in January 2014, no filling level data was measured because the whole experiment was shut down completely for the installation of the UPS power lines. The dewar vessel is refilled bi-weekly by a service provider on site. 

\subsection{Preamplifier cooling}
The preamplifiers produce approximately \SI{30}{W} of waste heat. This heat cannot dissipate passively and results in a higher temperature of those devices which deteriorates their signal 
quality. To prevent this warming, and to even cool the devices below ambient temperature, a cooling system was installed. Metal plates that are being flushed with cooled water are placed above all preamplifier devices (i.e. in between them as they are placed on top of each other). The 
water cooling system Julabo FL 601 is placed outside the building so that the waste heat does not affect the experiment. The actual working condition of the cooling system can be controlled remotely.

\subsection{Pulse generator}
A Berkeley Nucleonics PB-5 Precision Pulse Generator is installed, which can inject defined electric pulses into each of the 128 preamplifier channels at a time. This device is used for two main reasons.
One is to monitor the functionality and long-term stability of the electronic readout system. Therefore, a sequence of defined pulses is used in each data-taking run.
The other reason is to inject pulses to synchronize the FADCs. This timing precision is needed for coincidence analysis, which is used to identify events happening at the same time in several detectors.
The pulse generator is located on the second floor. The signals are transmitted differentially directly to the preamplifier devices, where they are converted to single-ended signals. All pulse generator pulses are flagged as such in the meta data by the FADCs, so that they can be discriminated against pulses from physics events.

\subsection{Calibration}
\begin{figure}
 \includegraphics[width=0.99\columnwidth]{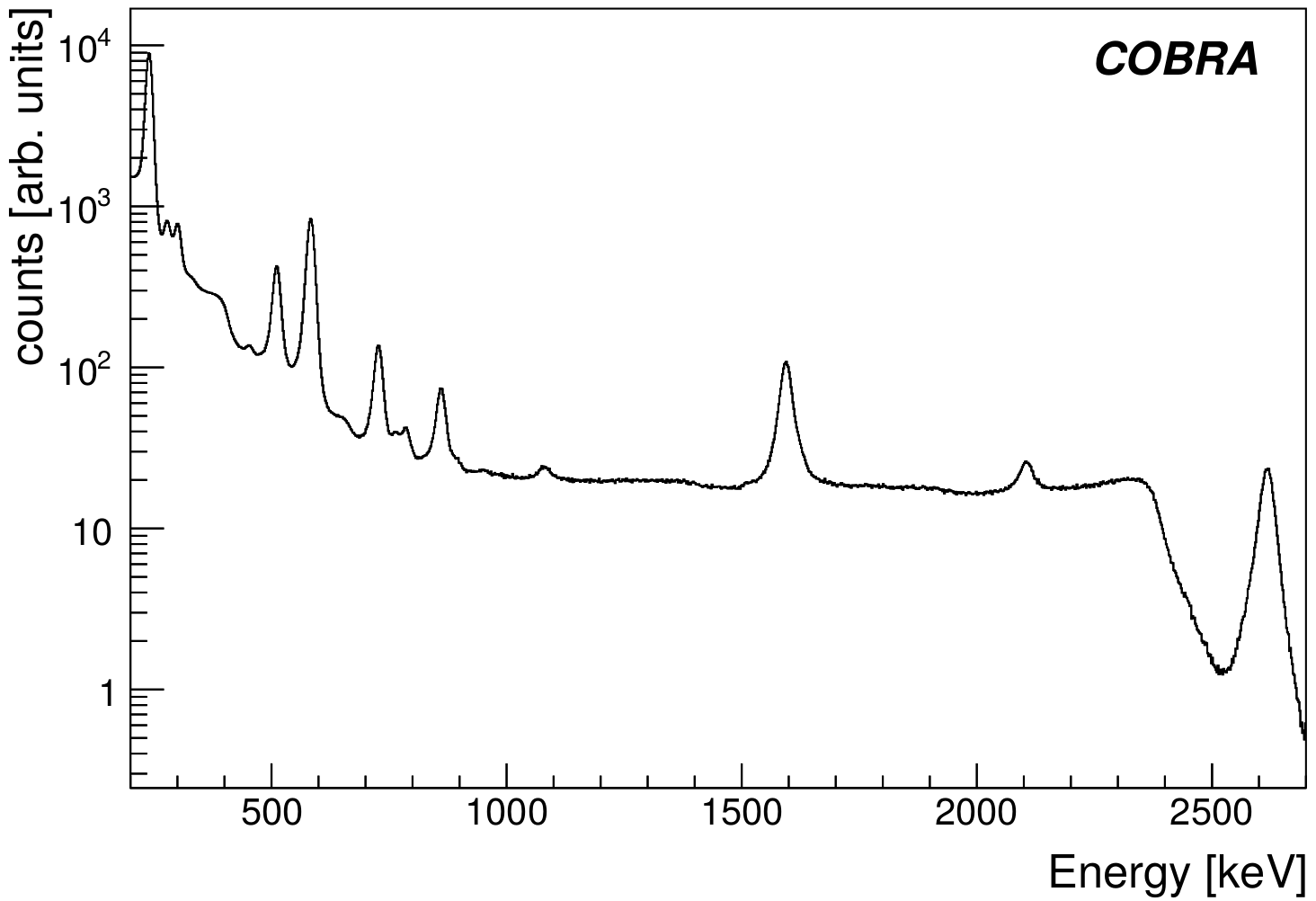}
 \caption[Calibration spectrum LNGS]{Exposure-weighted \isotope[228]Th calibration spectrum collected with the COBRA demonstrator.}
 \label{fig:calibration}
\end{figure}

Radioactive wire sources of \isotope[228]Th and \isotope[22]Na are used regularly to calibrate the detectors. Five Teflon tubes run through the shielding layers to the detectors, so that the calibration sources can be placed in the center above and below all four detector layers.
All tubes are arranged in a curved way, so that the shielding layers are not perforated in a straight line pointing directly to the detectors.
Each data-taking period has its own pre- and post calibration at the beginning and end, respectively. An example of a spectrum taken with a \isotope[228]Th source is shown in \autoref{fig:calibration}. The contributions from each detector are weighted according to their exposure collected during the physics runs.

\subsection{DAQ-server}
The main computer used for the experiment is located on the second floor. It can be accessed from outside via a network, and it provides all main services, like running the data acquisition and communication to all sub-systems.
The typical duration of each data-taking run is 4\,h. The data is stored in the \textsc{root} file format \cite{cern_root} and is copied and backed up regularly to external servers.

\subsection{Monitor system}
A set of sensors are installed to monitor and log the environmental conditions of the experiment.
Several sensors measure the temperature, humidity and pressure in the ground floor of the building, within the outer shielding on top of the preamplifiers, and in the inner shielding on top of the lead bricks. 
In particular, the preamplifier cooling and nitrogen-flushing are monitored.
Furthermore, the preamplifier's supply voltage is monitored, as well as the filling level and temperature of the nitrogen dewar vessel. Several cameras allow for real-time images of the setup.

\section{Performance of the demonstrator}
\label{sec:performance}
\begin{figure}
 \includegraphics[width=0.99\columnwidth]{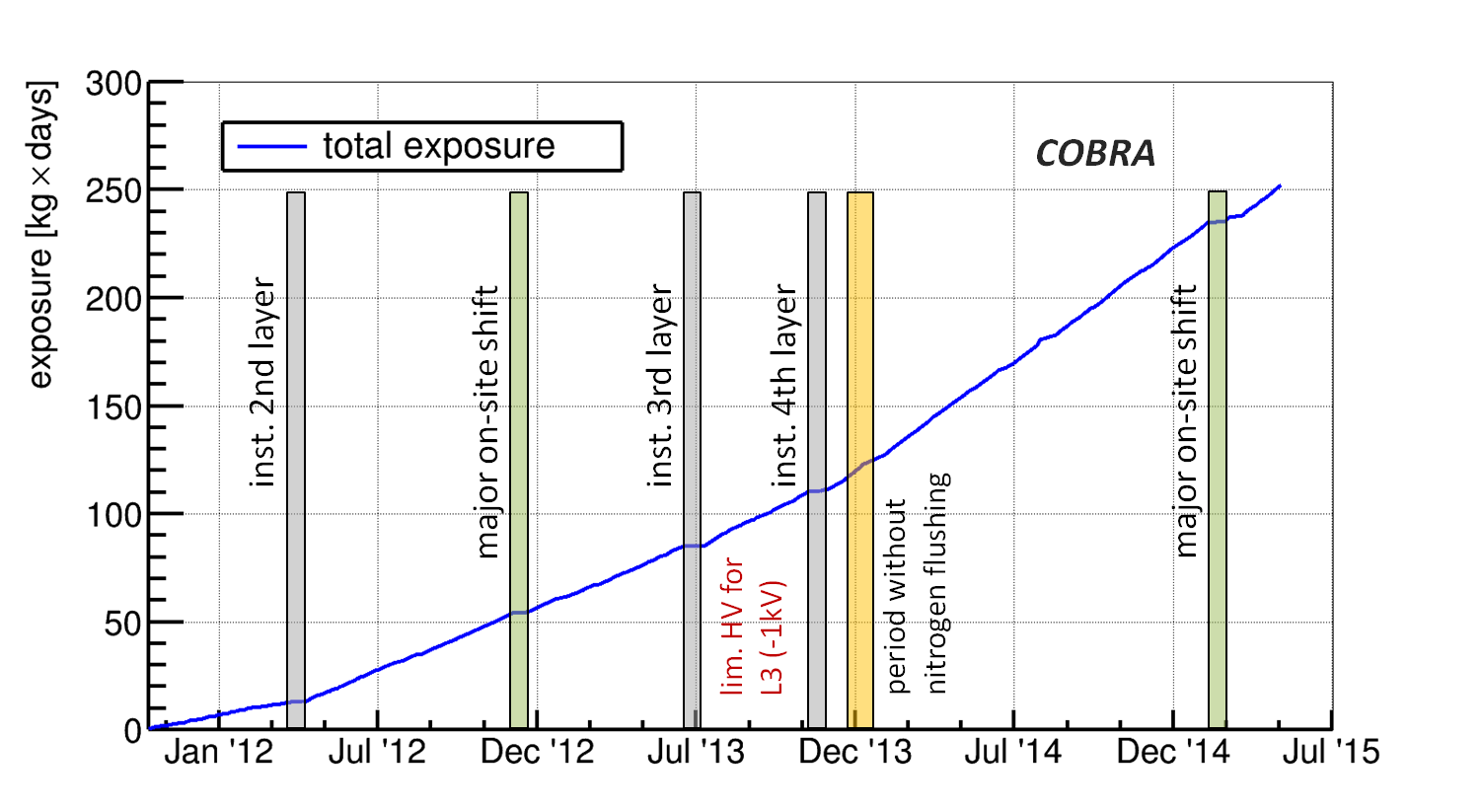}
 \caption[Exposure]{Exposure of the COBRA experiment, indicated are the upgrade works on site, notable periods where the nitrogen-flushing failed and detector layer 3 had to be operated with lower bias voltage. In the beginning of 2016 1\,kg\,y can be achieved.}
 \label{fig:exposure}
\end{figure}

Currently 61 of the 64 detectors are functional, which corresponds to a yield of more than 95\%. Two of those three detectors suffer from faulty voltage contacts.\\
Under normal data-taking conditions, i.e. without periods of larger external power cuts, the demonstrator's life-time is only reduced by calibration measurements. These take typically one day per month, so the data taking efficiency is above 95\%.
To date, more than 250 kg\,d of high quality low-background physics data have been collected, which are used for physics analysis.
\autoref{fig:exposure} shows the collected exposure as a function of time. Periods without data-taking are indicated.\\
The average count rate of the COBRA demonstrator above 500\,keV is 225 counts/day, corresponding to a rate of 3.6 counts/(day$\cdot$detector), see \autoref{fig:nitrogen_flushing}. In comparison, if one such detector is operated unshielded on the earth's surface under similar conditions, the rate is approximately $2\cdot10^4$ counts/day. 
This corresponds to a reduction of four orders of magnitude and is a result of operation in an underground laboratory, the shielding concept and selecting low-background materials.\\
Auxiliary measurements are performed to quantify the quality of the different components of the readout chain. A pulse generator injects pulses to the FADCs directly, then the pulses are converted to differential signals and vice versa before being measured by the FADCs. 
Finally, the whole DAQ chain of preamplifier, differential signaling, linear amplifier and FADC is tested. The amplitudes of the pulse heights are filled into a histogram. The full width at half maximum (FWHM) of the resulting distribution is a measure for the resolution of the devices under test. The resolution of the FADC has a FWHM of 0.05\%. The conversion to differential signaling and vice versa results in a resolution of 0.1\%.
The resolution of the preamplifier devices cannot be measured directly, only in conjunction with the whole DAQ chain. The total resolution varies between 0.4\%\,FWHM and 1.0\%\,FWHM due to the variation of the different components used in the different channels. 
The resolution of the preamplifier channels can hence be estimated to be between 0.3\% and 0.9\%.\\
As can be seen in \autoref{fig:calibration}, the energy resolution of the \isotope[208]Tl peak at 2\,615\,keV is 1.4\% FWHM. The best detector has a resolution of 0.8\%.

\section{Summary and outlook}
\label{sec:summary}The COBRA collaboration searches for neutrinoless double beta-decay with a focus on \isotope[116]Cd. A demonstrator setup was built in the Gran Sasso underground laboratory between 2011 and November 2013. It consists of 64 CdZnTe detectors operated in a coplanar-grid mode.
The detector, shielding concept, data acquisition system and experimental infrastructure have been optimized to meet the requirements of low-background physics experiments. 
After several working shifts at the LNGS, the demonstrator operates stably and is taking data reliably. 
To date, more than 250\,kg\,d of high-quality low-back\-ground physics data have been collected. The data-taking will continue to gain more exposure for physics analyses and to monitor the long-term stability of the detectors.\\
To be able to measure the long half-life of neutrinoless double beta-decay, the demonstrator has to be extended to a large-scale experiment by a factor of about 1\,000.	
Currently, the COBRA collaboration is preparing to build one detector module, the smallest unit of such a large-scale experiment. This module will consist of nine detectors with a volume six times larger than the ones which are currently being operated.
An ASIC-based integrated readout system will be used, which will improve the energy resolution, reduce the power consumption and allow easy scaling to larger detector numbers. The integration of that new detector module into the COBRA demonstrator at the LNGS is planned for 2016, together with a repair of the faulty voltage contacts of the two non-functional detectors.
With this new detector module, the rate with which the exposure will be collected during the following years will increase by a factor of about two.

\section*{Acknowledgement}
We thank the LNGS for the continuous support of the COBRA experiment. We explicitly thank Ralf Wischnewski, Hardy Simgen, Matthias Laubenstein and Matt\-hias Junker for their support and contributions to the experiment. COBRA is supported by the German Research Foundation DFG.

% \section*{References}

\bibliography{cobra_demonstrator}
\bibliographystyle{elsarticle-num}

\end{document}